\documentclass[useAMS,usenatbib,referee]{mn2e}
\usepackage{rotating,epstopdf}
\usepackage{graphicx}
\usepackage{color}

\title[Search for the Gamma-ray Emission from M33 with Fermi LAT]
  {Search for the Gamma-ray Emission from M33 with Fermi LAT}
\author[Lei Fu et al.]
{L.~Fu$^{1}$, Z.~Q.~Xia$^{1,2}$, Z.~Q.~Shen$^{1,3}$.\\
  $^{1}$Key Laboratory of Dark Matter and Space Astronomy, Purple Mountain Observatory, Chinese Academy of Sciences, Nanjing 210008, China\\
  $^{2}$School of Physics, University of Science and Technology of China, Hefei, 230026, China\\
  $^{3}$University of Chinese Academy of Sciences, Yuquan Road 19, Beijing, 100049, China}
\begin{document}

\label{firstpage}
\maketitle

\begin{abstract}
In the work we search for the $\gamma$-ray signal from M33, one of the biggest galaxies in the Local Group, by using the Pass 8 data of Fermi Large Area Telescope (LAT). No statistically significant gamma-ray emission has been detected in the direction of M33 and we report a new upper limit of high energy ($>100\,\rm MeV$) photon flux of $2.3\times 10^{-9}\,\rm ph\,cm^{-2}\,s^{-1}$, which is more strict than previous constrains and implies a cosmic ray density of M33 lower than that speculated previously. Nevertheless the current limit is still in agreement with the correlation of star formation rate and $\gamma$-ray luminosity inferred from the Local group galaxies and a few nearby starburst galaxies.
\end{abstract}

\begin{keywords}
Gamma rays: observations --- galaxies: individual (M33)

\end{keywords}

\section{INTRODUCTION}
Since cosmic rays (CRs) produce $\gamma$-ray photons through interactions with the interstellar medium and radiation fields \citep{smp2007}, the diffused $\gamma$-ray emission as an important indirect measurement of the CRs propagation and interstellar medium distribution in galactic environment, has received more and more attention in recent decades.
However, on the one hand the high energy $\gamma$-rays are much rare than for example the optical photons and the effective area of space instruments can not be very large due to the tight constraint on the power/weight of the detector as well as the limited budget. On the other hand the characterization of high-energy photons are difficult (e.g. the angular resolution is only degrees to arcminutes at best, being a function of energy). Thus the catalogs of gamma-ray sources have long been occupied with unidentified labels, e.g. in the Second COS-B catalog more than $80\%$ sources were unidentified \citep{sbb+1981} and in the Third EGRET Catalog this number is nearly $60\%$ \citep{hbb+1999}. Such a puzzling situation has changed dramatically in 2008 thanks to the successful launch/performance of the {\it Fermi} Gamma-ray Space Telescope \citep{aaa+2009a}. With an unprecedented large $\sim 8000~{\rm cm}^2$ effective area, a wide  field of view ($\sim 2.3$ sr) and a short dead time ($>95\%$ duty cycle) together with unprecedent sensitivity from $\sim 20$ MeV to $> 300$ GeV range \citep{aaa+2009a}, all these made the Large Area Telescope (LAT) on board {\it Fermi} a great detector that can revolutionize our knowledge of the $\gamma-$ray sky.

The galaxies in our Local Group, due to their proximity, are excellent targets for $\gamma$-ray observations. Using the data of LAT, \citep{aaa+2009b} reported the detection of diffused $\gamma$-ray emission of the Galactic plane in the energy ranges from 100 MeV to 10 GeV, unlike the results of EGRET displaying a GeV excess \citep{hbc+1997} the ``new" spectrum is consistent with a diffuse Galactic $\gamma$-ray emission model \citep[see][for the details of the model]{aaa+2009b}. Soon after that the Large Magellanic Cloud (LMC), the Small Magellanic Cloud (SMC) and the Andromeda (M31) in our Local Group have also been detected by the LAT. The LMC has been detected at a 33$\sigma$ significance level with the first 11 months data of LAT, and the integrated photon flux $>100{\rm MeV}$ \citep{aaa+2010a} is compatible with the previous EGRET observations \citep{sbd+1992}. More recent LAT data analysis of LMC can be found in \citet{aaa+2016}. The LAT results of the $>100\,{\rm MeV}$ flux from the SMC \citep{aaa+2010b} is consistent with the upper limit derived from the EGRET data \citep{sbd+1993}. For M31, since its relatively distant \citep[about $800\,{\rm kpc}$, see][]{Holland1998,Walker2003} compared with a distance only $50-60\,{\rm kpc}$ of the LMC and SMC, the LAT observation in the first two years yields a detection of $5.3\sigma$ significance level in the energy range of $200\,{\rm MeV}-20\,{\rm GeV}$ \citep{aaa+2010c}. The gamma-ray emission of galaxies in our Local Group has also been investigated in \citet{lw2011}.

As the third largest galaxy in our Local Group, M33 has been considered to be another promising gamma-ray source due to its high mass and proximity. Before {\it Fermi} era, COS-B \citep{pmb+1981} and EGRET \citep{hbb+1999} both  searched for the high energy $\gamma$-ray photons from M33 but only yielded upper limits. The upper limit of $>100\,{\rm MeV}$ flux derived by EGRET is $3.6\times 10^{-8}\,{\rm ph\,cm^{-2}\,s^{-1}}$. By using a nearly 2 years accumulating data of LAT, \citet{aaa+2010c} also  searched for the $\gamma$-ray emission from M33. Again, no significance gamma-ray emission was detected and a more stringent upper limit of $>100\,{\rm MeV}$ flux of $5.1\times 10^{-9}\,{\rm ph\,cm^{-2}\,s^{-1}}$ was derived.

Since then the {\it Fermi} team has released three update version of data, from the Pass6, Pass7, Pass7 reprocessed to the latest Pass8 release.The latest Pass 8 data release has several very important improvements, including an extension of the energy range, better energy measurements and a larger effective area etc \citep{aab+2013}. Benefited from these remarkable improvements, with the Pass 8 data it may be possible to detect the significant gamma-ray emission in the direction of M33. Motivated by such a possibility in this work we analyze the 7.5 years Fermi-LAT pass 8 data. This work is arranged as the following: in Section 2 we introduce the data selection, then we present the data analysis and results in Section 3, in Section 4 we summarize our results with some discussions.

\section{Data analysis}
\subsection{Data selection}

We use 88 months of Pass 8 LAT data from 2008 Oct 27 (MET = 246823875s) to 2016 Mar 01 (MET = 478483204s). Photons with energy range between 300 MeV and 300 GeV are taken into consideration. We use the SOURCE event class with the standard conversion-type (FRONT+BACK) selection, recommended for point source analysis. And we apply the zenith angle cut ${\rm \theta} < 90^{\rm \circ}$ to reduce the contribution from Earth Limb. Fermi Science Tools v10r0p5 and instrument response functions (IRFs) P8R2\_SOURCE\_V6 are used for this analysis, which are available from the Fermi Science Support Center\footnote{http://fermi.gsfc.nasa.gov/ssc/}.

\subsection{Data analysis}

\begin{figure}
\includegraphics[width=1\columnwidth]{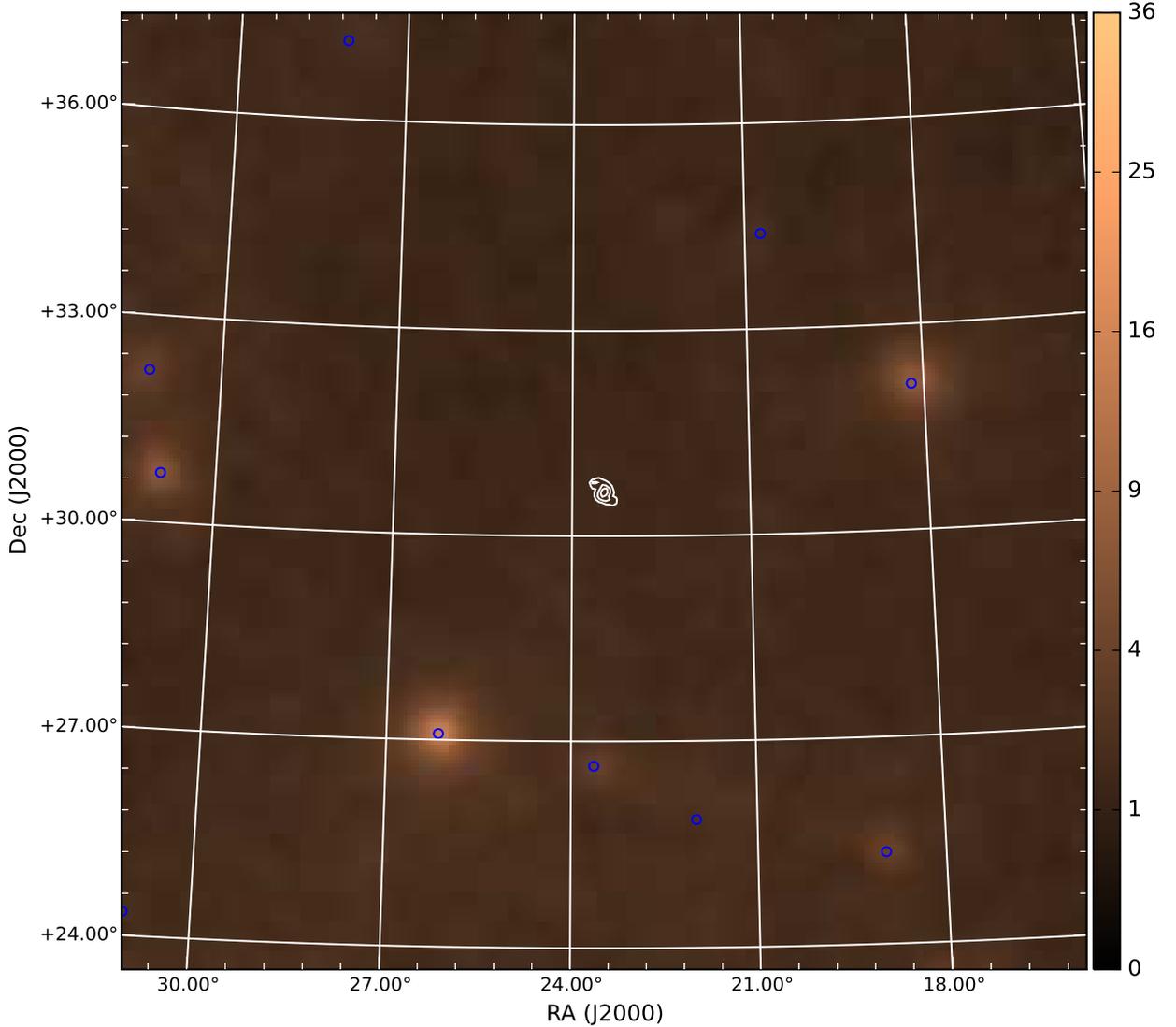}
\caption{A gaussian kernel (${\rm \sigma} = 0.15^{\rm \circ}$) smoothed counts map of the ROI $14^{\rm \circ} \times 14^{\rm \circ}$ from 300 MeV to 300 GeV with a pixel size of $0.05^{\rm \circ} \times 0.05^{\rm \circ}$ together with the positions of sources (blue dot) listed in the 3FGL catalog. The white contours are IRIS 100 ${\rm \mu m}$ contours of M33.}
\label{counts map}
\end{figure}

We create $14^{\rm \circ} \times 14^{\rm \circ}$ regions of interest (ROI) centered on (RA, DEC, J2000) = ($23.4621^{\rm \circ}$, $30.6599^{\rm \circ}$) and perform a standard binned likelihood analysis with $0.05^{\rm \circ}$ spatial bins and 30 logarithmic energy bins. A counts map of the ROI together with the positions of sources listed in the 3FGL catalog \citep{aaa+2015} are shown in Fig.\ref{counts map}. Overlaid are IRIS 100 ${\rm \mu m}$ contours of M33.

To look for the signal from M33, we perform maximum likelihood ratio tests. The Test Statistic (TS) is defined as ${\rm TS = -2ln(}L_{\rm max,0}/L_{\rm max,1})$ following \citet{mbc+1996}, where $L_{\rm max,0}$ is the maximum likelihood value for a model without an additional source (the null hypothesis) and $L_{\rm max,1}$ is for a model with that additional source(the alternative hypothesis).

The background is modelled as a combination of diffuse backgrounds and all point sources from the 3FGL catalog found within a $20^{\rm \circ}$  radius around the M33. To model the diffuse backgrounds, we use the templates for the Galactic diffuse emission (gll\_iem\_v06.fits) and an isotropic component (iso\_P8R2\_SOURCE\_V6\_v06.txt), which are provided by the Fermi-LAT collaboration. We free the normalization and spectral index of all sources within $7^{\rm \circ}$  from the target source in a maximum likelihood fit. We also let free the flux normalization of the Galactic diffuse and isotropic components.

The M33 spatial template is derived from the Improved Reprocessing of the IRAS Survey (IRIS) 100 ${\rm \mu m} $ far infrared map \citet{mdl2005}. The majority of gamma-ray emission from a galaxy is due to interactions of high-energy cosmic rays with interstellar gas and radiation fields in the galaxy. Therefore, far infrared emission, which traces interstellar gas convolved with the recent massive star formation activity, is a reasonable approximation for the expected distribution of gamma-ray emission from a galaxy to the first order \citep{aaa+2010c}.In order to obtain the M33 spatial template, we set any pixel less than 5 ${\rm MJy/sr}$ to 0 in the IRIS 100 ${\rm \mu m} $ map to remove the background emission and normalize this map. We use this IRIS 100 ${\rm \mu m} $ spatial template and a power-law spectral shape to perform this fit using gtlike, and obtain TS=17.26 which does not reach the detection significance of $5{\rm \sigma}$ for 2 free parameters.

\begin{figure}
\includegraphics[width=1\columnwidth]{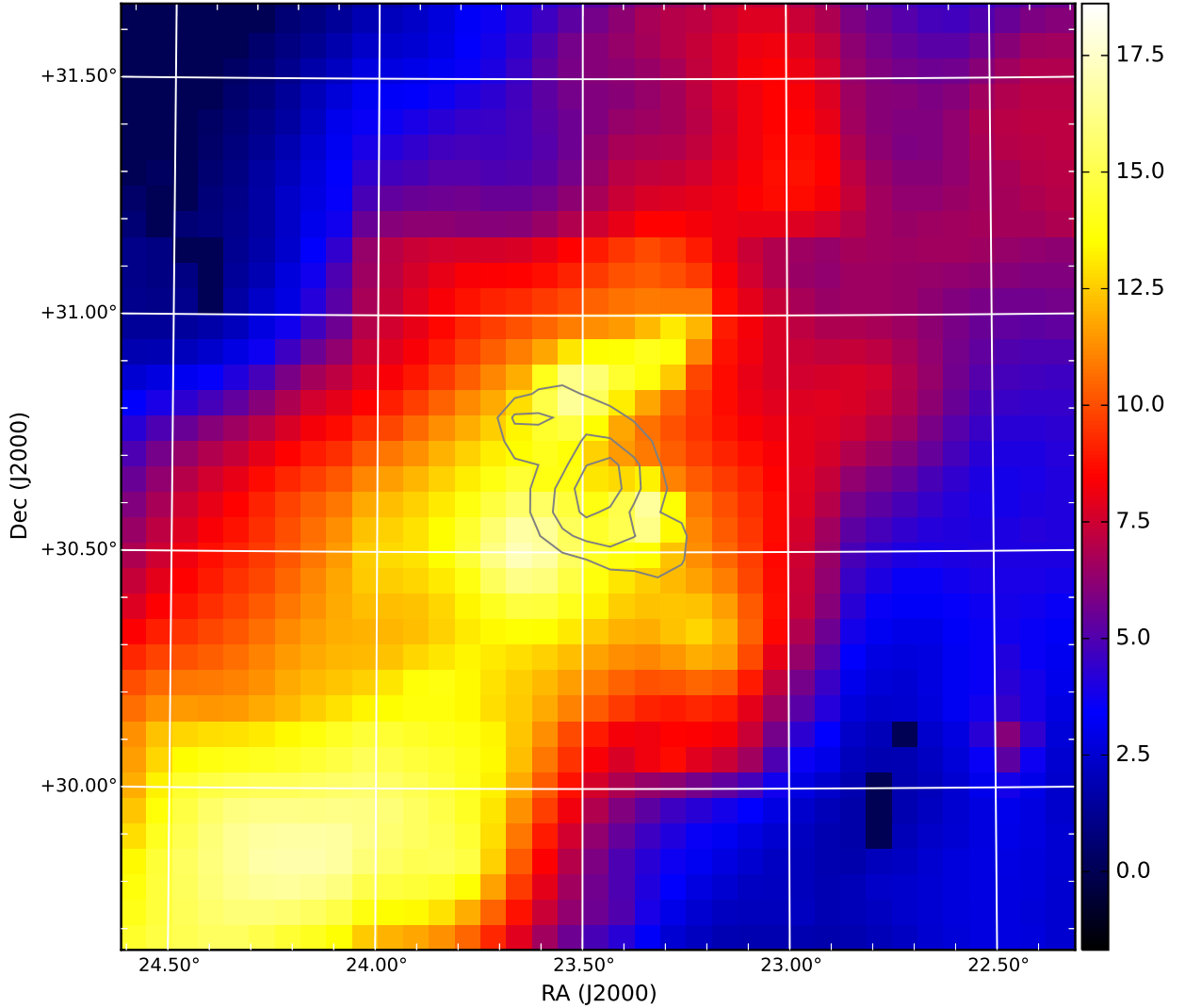}
\caption{$2^{\rm \circ} \times 2^{\rm \circ}$ TS map centered at the nominal position of M33. The grey contours are IRIS 100 ${\rm \mu m} $ contours of M33}
\label{Tsmap}
\end{figure}

Then we derive  a $2^{\circ} \times 2^{\circ}$ TS map by placing a test point source at the location of each pixel of the map and maximizing the likelihood which is implemeted in the gttsmap tool\footnote{http://fermi.gsfc.nasa.gov/ssc/data/analysis/scitools/help/gttsmap.txt}. The location with the maximum TS is the best fit source position. The location of ${\rm TS_{max}}$  ($23.66^{\rm \circ}$, $30.48^{\rm \circ}$, J2000) is $0.27^{\rm \circ}$ away from the center. And the ${\rm TS_{max}}$ = 17.11  is not enough to say that the source on this location is real.  As we can see from Fig. \ref{Tsmap}, the TS map is inconsistent with M33.

To find whether there is a flare, we compute TS maps separately in 10 time bins . The result is irregular and many of them don't have any signal within the center of ROI. Then we subdivide the time bins with the ${\rm TS_{max}}$ $>$16, however we don't find any flare with TS $>$ 25 that corresponds to a local significance of $5{\rm \sigma}$. So we conservatively suggest that the TS obtained from the above analysis is a result of statistical fluctuations and systematic errors including uncertainties in the performance of the LAT and uncertainties in diffuse Galactic model.

Without any evidence for the emission from M33, we conservatively do the upper limit analysis for M33. In order to construct the spectral model, we multiply the flux in each pixel of Fermi Galactic diffuse emission template (gll\_iem\_v06.fits) by the corresponding solid angle and sum them up. We  derive an 95\% confidence flux upper limit of ${\rm 2.3\times10^{-9}~ph~cm^{-2} s^{-1}}$ in the 100 MeV$-$300 GeV energy range using this Galactic diffuse spectral model and the IRIS 100 ${\rm \mu m}$ spatial template, which is stronger than \citet{aaa+2010c}\footnote{During the revision of this manuscript, we notice that Fermi group published their latest results of the gamma-ray signal from M33 which give only the upper limit and is in well agreement with our results\citep{aaa+2017}.}.

Considering there is a low significance excess of gamma-ray emission observed with TS$\sim$17, we want to estimate the expected sensitivity of M33. We run Monte Carlo simulations to generate 100 random observation scenarios for the background-only hypothesis and calculate the 95\% confidence expected flux upper limits for M33 in these observation simulation scenarios. The average of these expected flux upper limits in the background-only hypothesis is ${\rm 7.42\times10^{-10}~ph~cm^{-2} s^{-1}}$.

\section{Discussion}

\begin{figure}
\includegraphics[width=\columnwidth]{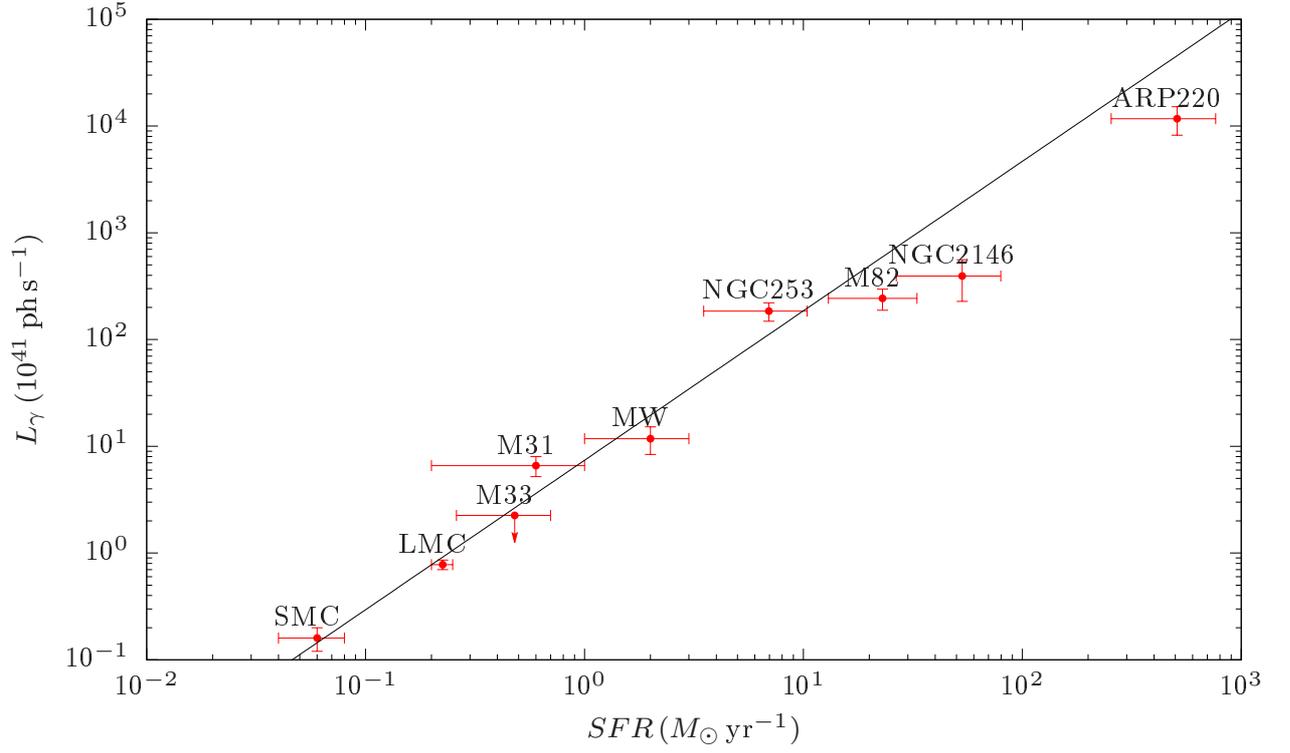}
 \caption{$>100\,\rm MeV$ $\gamma$-ray luminosity vesus star formation rate. We also include two nearby starburst galaxies M82 and NGC253. The solid line is the power law fits for the MW, M31 and Magellanic Clouds.}
 \label{lsfr}
\end{figure}

Usually the $\gamma$-ray emissivity of a galaxy can be described/characterized by the ratio between $\gamma$-ray luminosity and total hydrogen atom number of the galaxy (${\bar q}_{\gamma}=L_{\gamma}/N_{\rm H}$). Following \citet{aaa+2010c} in Table \ref{table1} we calculate ${\bar q}_{\gamma}$ of M33 and other galaxies in our Local Group but with the more recent observation data. Also we include four nearby star burst galaxies M82, NGC253, ARP220 and NGC 2146. The $>100\,\rm MeV$ photon luminosities are $L_{\gamma}=4\pi d^2 F_{\gamma}$, where $d$ is the distance of the galaxy. The total number of hydrogen atoms $N_{\rm H}=1.19\times 10^{57}M_{\rm H}$, in which the mass of hydrogen atoms $M_{\rm H}$ is in units of $M_\odot$ and includes the masses of HI ($M_{\rm H}$) and $\rm H_{2}$ ($M_{\rm H_2}$). The quoted errors of $L_{\gamma}$ and ${\bar q}_{\gamma}$ include that of $d$ and $M_{\rm H}$. The derived $L_{\gamma}$ and SFR relation using the data of LMC, SMC, MW and M31, i.e.,$L_{\gamma}=(7.4\pm1.6)\times 10^{41} \rm \, ph\,s^{-1} ({\rm SFR}/M_{\odot}\,\rm yr^{-1})^{1.4\pm0.3}$ which reported in \citet{aaa+2010c} is plotted in Fig.\ref{lsfr}.

We find that the relation are in well agreement with M33 (the upper limit) and the nearby four starburst galaxies M82, NGC253, ARP220 and NGC2146, suggesting that the tight correlation of $L_{\gamma}-{\rm SFR}$ holds not only for the Local Group but also for starburst galaxies, which is consistent with the results found by \citet{aaa+2010c}. However such a conclusion is based on a very limited sample and we can not draw a universal conclusion. Further observation and more sample are crucially needed. If this relation exists, it will be similar to the well-known radio and far-infrared correlation which robust for a multiple types of galaxies, the explanation for the latter may be linked with the CR electron calorimetry \citep{aaa+2010c}. Moreover, the new upper limit of M33 correponding a luminosity of $L_{\rm 100\,Mev-300\,GeV}=1.6 \times 10^{38} \rm erg\;s^{-1}$ which is consistent with the fit uncertainty of the scaling relations between tracers of SFR (1.4 GHz continuum, total IR) and gamma-ray luminosity given by \citet{aaa+2012}.

\begin{figure}
\includegraphics[width=\columnwidth]{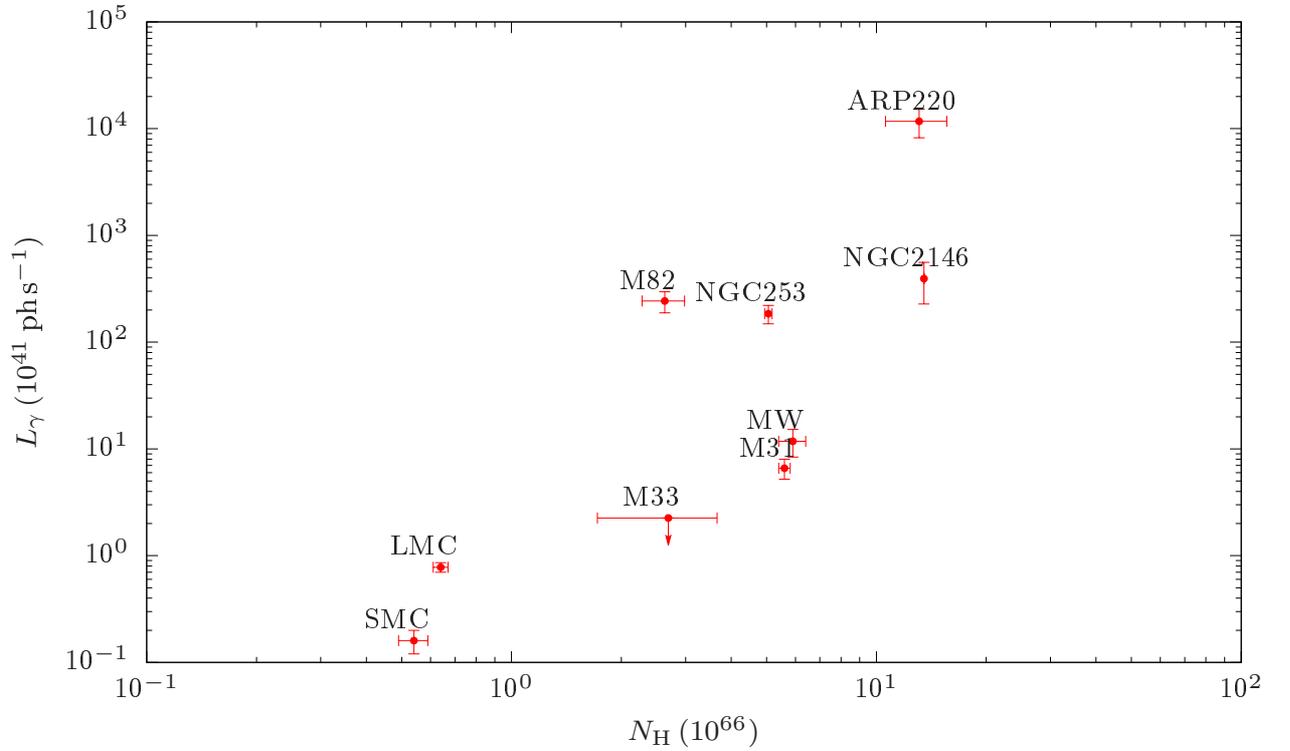}
 \caption{$>100\,\rm MeV$ $\gamma$-ray luminosity vesus the total number of hydrogen atoms.}
 \label{lnh}
\end{figure}

The inelastic collision between energetic CR nuclei and interstellar gas with energies above anti-coincidence dome can produce neutral pion which with a possibility of 99\% decay into two $\gamma$-ray photons in a very short lifetime. This process is believed to dominate the generation of $\gamma$-ray photons with energies $>100\,\rm MeV$ \citep[see e.g.][]{sbd+1998,pf2001}. Since interstellar gas, CRs, supernovae and star formation are connected to some extent, many authors have argued that galaxy $\gamma$-ray luminosities and SFRs are correlated \citep[e.g.][]{pf2002,tks+2004,tqw2007,stecker2007,pr2010,ltq+2011}. Thus the ${\bar q}_\gamma$ value reflects the CR density of a galaxy, and the ratio of ${\bar q}_\gamma$ between different galaxies evaluate the relative average CR density of them. This value differs from galaxies with different SFR, supernova rate, interstellar medium etc. Fig.\ref{lnh} is the $N_{\rm H}$ vs $L_{\gamma}$ plot. The upper limit CR density of M33 is about half of MW.

The relative distant, low SFR and hydrogen mass properties of M33 makes its detection much more challenging in comparison with other gamma-ray detected galaxies in our Local Group. \citet{pf2001} predicted the possibility of M33 detectable for LAT by assuming that the CR flux is proportional to the supernova rate. They found that the predicted flux of M33 is $\sim \rm 1.1\times 10^{-9}~ph~cm^{-2} s^{-1}$ with energies $>100\,\rm MeV$, which is slightly below the sensitivity limit of LAT ($\sim \rm 2\times 10^{-9}~ph~cm^{-2} s^{-1}$), and M33 will be detected by LAT with a 4.1$\sigma$ significance after 10 years of observation. Using a scaling relation of $\gamma$-ray luminosity and other photometric tracers of star formation (e.g. infrared, radio luminosity). \citet{aaa+2012} predicted a much lower chance of detecting the gamma-ray emission from M33 by LAT in 10 years observation and the significance is expected to be 2.1$\sigma$. Therefore the non-detection of the gamma-ray emission from M33 seems to be well consistent with the theoretical prediction. Indeed our latest upper limit of $>100\,\rm MeV$ flux from M33$\sim {\rm 2.3\times10^{-9}~ph~cm^{-2} s^{-1}}$, which is very similar to the current point source sensitivity of LAT 10 years observation \footnote{http://www.slac.stanford.edu/exp/glast/groups/canda/lat\_Performance.htm}. The prospect of significantly detecting the gamma-ray emission from M33 is thus still uncertain.

\begin{table*}
\centering
\begin{center}
\caption{Parameters and gamma-ray flux of Local Group and nearby starburst galaxies.}
\begin{tabular}{lccccccc}
\hline\hline
Galaxy	 & Distance	 & $M_{\rm HI}$	  & $M_{\rm H_{2}}$	 & {\it SFR}			 & $F_{\gamma}$					 & $L_{\gamma}$			 & $\bar{q}_{\gamma}$ \\
& kpc	 & $10^8 M_{\odot}$ & $10^8 M_{\odot}$	 & $M_{\odot}\,{\rm yr^{-1}}$	 & $10^{-8}\,{\rm ph\,cm^{-2}\,s^{-1}}$	 & $10^{41}\,{\rm ph\,s^{-1}}$ & $10^{-25}\,{\rm ph\,s^{-1}\,H}$-atom$^{-1}$ \\
\hline
 MW  & \ldots        & $35\pm4^{(1)}$   & $14\pm2^{(2)}$ & $1-3^{(3)}$ & \ldots & $11.8\pm3.4^{(4)}$ & $2.0\pm0.6$ \\
 M31 & $780\pm33^{(5)}$ & $42.7^{(6)}$ & $3.6\pm1.8^{(7)}$ & $0.2-1^{(8)}$ & $0.9\pm0.2^{(9)}$ & $6.6\pm1.4^{(9)}$ & $1.17\pm0.28$  \\
 M33 & $847\pm60^{(10)}$ & $19\pm8^{(11)}$ & $3.3\pm0.4^{(11)}$ & $0.26-0.7^{(12)}$ & $<0.23$ & $<2.3$ & $<1.35$ \\
 LMC & $50\pm2^{(13)}$ & $4.8\pm0.2^{(14)}$ & $0.5\pm0.1^{(15)}$ & $0.20-0.25^{(16)}$ & $26.3\pm2.0^{(17)}$ & $0.78\pm0.08^{(9)}$ & $1.2\pm0.1$ \\
 SMC & $61\pm3^{(18)}$ & $4.2\pm0.4^{(19)}$ & $0.25\pm0.15^{(21)}$ & $0.04-0.08^{(22)}$ & $3.7\pm0.7^{(22)}$ & $0.16\pm0.04^{(9)}$ & $0.31\pm0.07$ \\
 M82 & $3630\pm340^{(23)}$ & $8.8\pm2.9^{(24)}$ & $13^{(25)}$ & $13-33^{(26)}$ & $1.54\pm0.19^{(27)}$ & $242.7\pm54.4$ & $92.4\pm24.1$\\
 NGC253 & $3500\pm200^{(28)}$ & $25^{(29)}$ & $17\pm1^{(30)}$ & $3.5-10.4^{(31)}$ & $1.26\pm0.2^{(27)}$ & $184.6\pm36.1$ & $36.5\pm7.2$\\
 ARP220& 74700$^{(32)}$ & $25\pm21^{(33)}$ & 84$^{(34)}$ & $254.8-764.3^{(35)}$ & $0.18\pm0.05^{(36)}$ & $11700\pm3500$ & $894.4\pm315.5$\\
 NGC2146& 15200$^{(32)}$ & $57^{(37)}$ & 55$^{(37)}$ & $26.6-79.7^{(35)}$ & $0.11\pm0.06^{(38)}$ & $393.9\pm165.8$ & $22.5\pm12.3$\\
\hline
\end{tabular}
\label{table1}
\end{center}
{\bf References.}
(1) \citet{pmg+2007}; (2) \citet{pmg+2007}; (3) \citet{yhp+2009}; (4) \citet{spd+2010}; (5) \citet{sg1998}; (6) \citet{ccf2009}; (7) \citet{nng+2006}; (8) \citet{yhp+2009,fgs+2013}; (9) \citet{aaa+2010c}; (10) \citet{gbf2004}; (11) \citet{gbr+2010}; (12) \citet{gbs+2007}; (13) \citet{ptg+2009}; (14) \citet{skc+2003}; (15) \citet{fkm+2008}; (16) \citet{hsk+2007}; (17) \citet{aaa+2010a}; (18) \citet{hhh2005}; (19) \citet{ssd+1999}; (20) \citet{lbs+2007}; (21) \citet{wkl+2004,blj+2011}; (22) \citet{aaa+2010b}; (23) \citet{kdg+2002}; (24) \citet{cly+2008}; (25) \citet{wws2002}; (26) \citet{fgl+2003}; (27) \citet{aaa+2012}; (28) \citet{rrm+2005}; (29) \citet{bof+2005}; (30) \citet{hwk+1997}; (31) \citet{lt2006}; (32) \citet{gs2004}; (33) \citet{bvs+1987}; (34) \citet{syb1997}; (35) \citet{ctr2005}; (36) \citet{pwl+2016}; (37) \citet{shl1995}; (38) \citet{twt2014}
\end{table*}

\section*{ACKNOWLEDGEMENTS}
We acknowledge the use of data from FSSC and SIMBAD database and thank the anonymous referee for the insightful comments/suggestions. We also thank Dr. Yizhong Fan for helpful discussion. This work was supported in part by 973 Program of China under grant 2013CB837000, the National Natural Science Foundation of China under grants 11273064, 11433009, 11573071, 11203009 and 11563003, and the Strategic Priority Research Program of CAS (under grant number XDB09000000).

\clearpage

\label{lastpage}
\end{document}